\pdfoutput=1 
\documentclass{JINST}
\usepackage{subfig}
\title{Development and Commissioning of the HARDROC based Readout for the INO-ICAL Experiment}

\author{Ashok Kumar\thanks{Corresponding author.}~, Ankit Gaur, Aman Phogat, Md. Rafik, Md. Naimuddin\\
\llap \noindent Department of Physics and Astrophysics\\
University of Delhi, Delhi, India - 110007\\
E-mail: \email{ashok.hep@gmail.com}}

\abstract{Glass based Resistive Plate Chambers (RPCs) are going to be used as an active element in the Iron Calorimeter (ICAL) experiment at the India based Neutrino Observatory (INO), which is being constructed to study atmospheric neutrinos. Though the RPC detector operational parameters are more or less finalized, the readout electronics is being developed using various technologies. The ICAL experiment will consist of about 29,000 RPC detectors of 2 m $\times$ 2 m in size with each detector having 64 readout channels both in the X and Y directions. The present study focusses on multi-channel electronics based on SiGe 350 nm technology as an option for the INO-ICAL RPC detectors. The study includes commissioning and usage of frontend application specific integrated circuit (ASIC) HARDROC chip in which 64 channels are handled independently to perform zero suppression. We present first testbench results using the HARDROC chip with the aim to use it finally in the ICAL experiment.}

\keywords{Resistive Plate Chamber; Iron Calorimeter; India based Neutrino Observatory; HARDROC}

\begin{document}

\section{Introduction}\label{sec:xxx}
Resistive Plate Chambers (RPCs) \cite{santonico} are gaseous detectors used for timing and triggering purposes in high energy experiments. Glass based RPCs are going to be used in the India-based Neutrino Observatory (INO) - Iron Calorimeter (ICAL) experiment for the detection of atmospheric neutrinos. This experiment can shed light on some of the unresolved mysteries encircling neutrino physics \cite{pramana, atmnu, epj}. RPCs are cheap and robust solutions for precise timing and triggering applications. In the previous R$\&$D studies \cite{ashok, nayeem} we have performed studies on 2 mm gap RPCs for their specific usage in the ICAL experiment. The performance of RPCs made up of two different resistive glasses, namely Saint Gobain and Asahi \cite{daljit, iworid}, has also been compared.

To cope with the tremendous number of electronic channels associated with this kind of detector we propose frontend electronics involving the HARDROC (HAdronic Rpc Detector ReadOut Chip) \cite{Callier}. The HARDROC is an application-specific integrated circuit (ASIC) based on SiGe 350 nm technology developed at AMS (Austrian Micro System). The readout is based on detector-embedded electronic boards equipped with daisy-chained 64 channel in order to minimize the number of output lines. With this semi-digital readout one can set 3 thresholds in order to extract good tracking information in the ICAL experiment. There is the possibility to use either the internal signal (OR of the 64 channels) or an external trigger. The second generation chip incorporates important features like analog storage, digitization and power pulsing. In this paper we introduce the HARDROC based readout for single gap RPCs and also present the new electronics and the readout system that was developed for this purpose. Finally, we describe the dedicated test-bench setup and first results obtained using this concept.

\begin{figure}
\centering
\includegraphics[height=10cm,width=14cm]{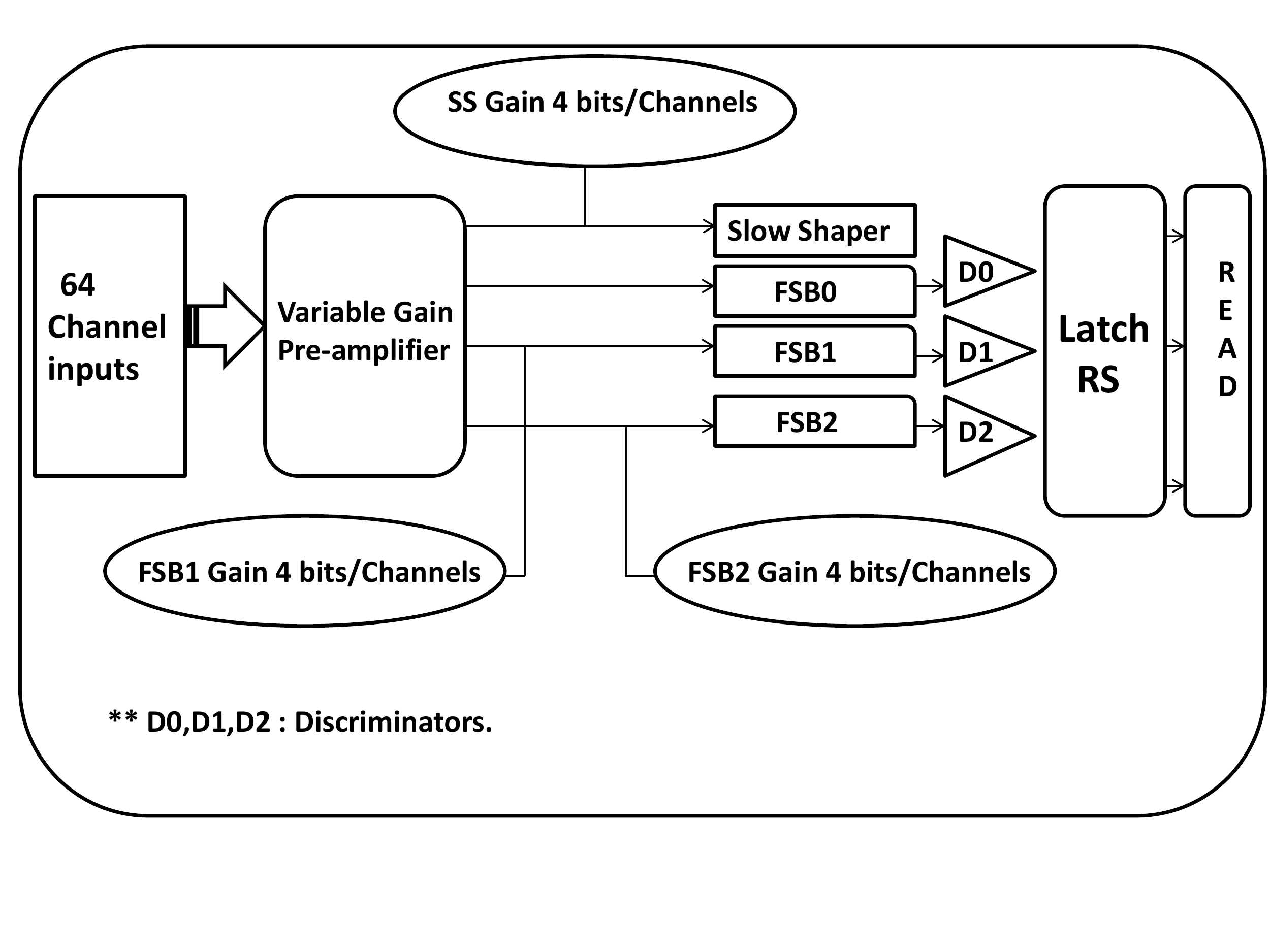}
\caption{Synoptic layout of the HARDROC chip.}
\label{layout}
\end{figure}

\section{Characteristics of HARDROC}\label{electrode}
The HARDROC ASIC is developed by the CALICE collaboration $\cite{Calice}$ to readout large area RPCs that would form a digital hadronic calorimeter at an International Linear Collider. A synoptic presentation of the HARDROC chip is shown in figure~\ref{layout}. Each input signal is first amplified using a variable gain preamplifier which exhibits low noise and low input impedance to minimize crosstalk. The gain can be tuned up to a factor of 2 to an accuracy of 6$\%$ with 8 bits/channel. The preamplifier is followed by 3 variable gain fast shapers and low offset discriminators. It has a variable slow shaper (50-150 ns) with track and hold feature to provide a multiplexed analog charge output up to 10 pC. This is a handle for diagnostic purposes. The thresholds are loaded by 3 internal 10 bit - DACs and the 3 discriminator outputs are sent to a 3 inputs to 2 outputs encoder. Each of these thresholds corresponds to an amount of integrated charge on the input. A 128 deep digital memory is provided to store the 2$\times$64 encoded outputs of the 3 discriminators. In order to avoid fake triggers produced by noisy channels, the output of each discriminator can be switched off from the trigger generator logic via the configuration parameters control (Slow Control hereafter) commands. Every 200 ns the status of the 64 lowest comparators is evaluated. If one of them is fired, the data are stored in an integrated digital memory, we call this mechanism "auto-triggering". Each channel can auto-trigger down to 4 fC without any external trigger or machine clock. The versatility of these chips allows their use in many applications including medical imaging, nuclear and astrophysics experiments. 

\section{HARDROC Testing and Performance}
The performance of the HARDROC has been measured using a testboard equipped with onboard software. In the chip on board (COB) version, the chip is bonded on the layer 5 of a 6 layers printed board. It provides the connection between adjacent ASICs and the FPGA ensuring the readout, control and external connectivity. The RPC signal mimicked using pulse generator is fed into the HARDROC. The signal output is analyzed using digital storage oscilloscope (DSO) and LabVIEW software installed on the computer. 

\begin{figure}
\centering
\includegraphics[height=5cm,width=14cm]{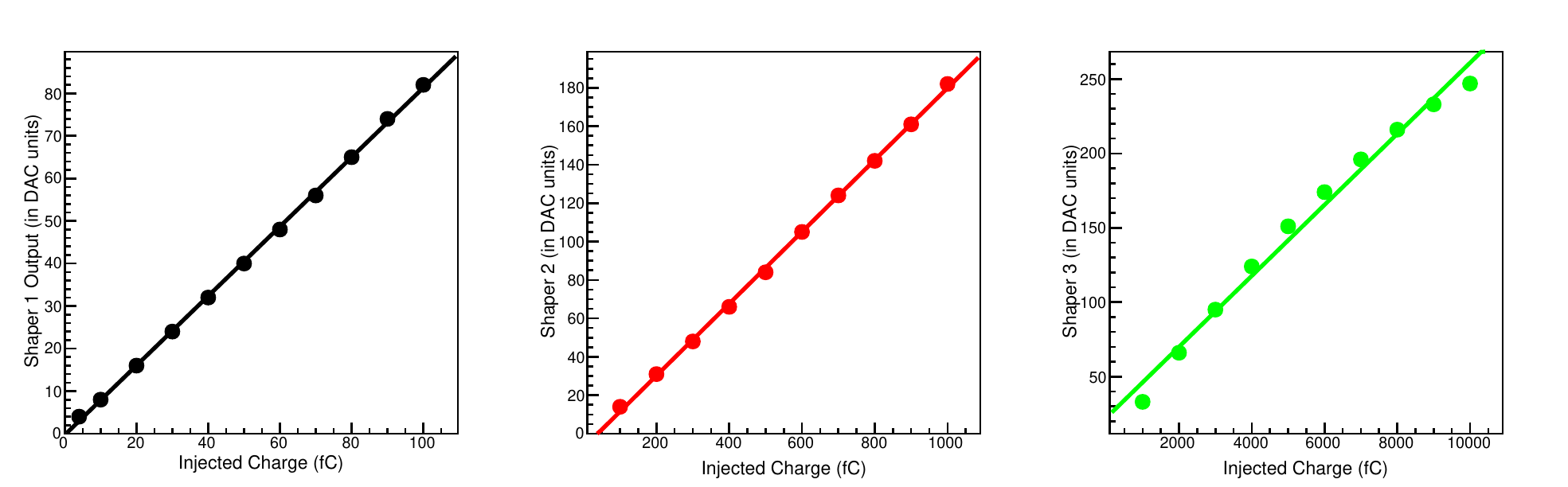}
\caption{The linear correlation between the DAC units and injected charge for all three shapers. In each shaper the input charge is injected as per the specified and allowed thresholds. The data is plotted at 50$\%$ efficiency.}
\label{ped}
\end{figure}

\subsection{Calibration Procedure}
We started with calibrating discriminator threshold for all the 64 channels. The test setup can handle 64 channels at a time with each of them having three independent comparators to provide three thresholds. Thresholds can be optimized through software to integer values in the interval between 0 and 1023 (DAC units). A charge input of 100 fC was injected in each channel for different gain settings. Each step was repeated many times to accumulate significant statistics. For each gain value the response of the studied channel was recorded and an efficiency curve (called S-curve) was obtained by varying the lower threshold by a step of 1 DAC unit. An automatic procedure using a fit is provided to determine the inflection point (corresponding approximately to 50$\%$ efficiency level) for each channel.

\begin{figure}[htbp]
\subfloat[]{\includegraphics[width=0.5\linewidth,height=5cm]{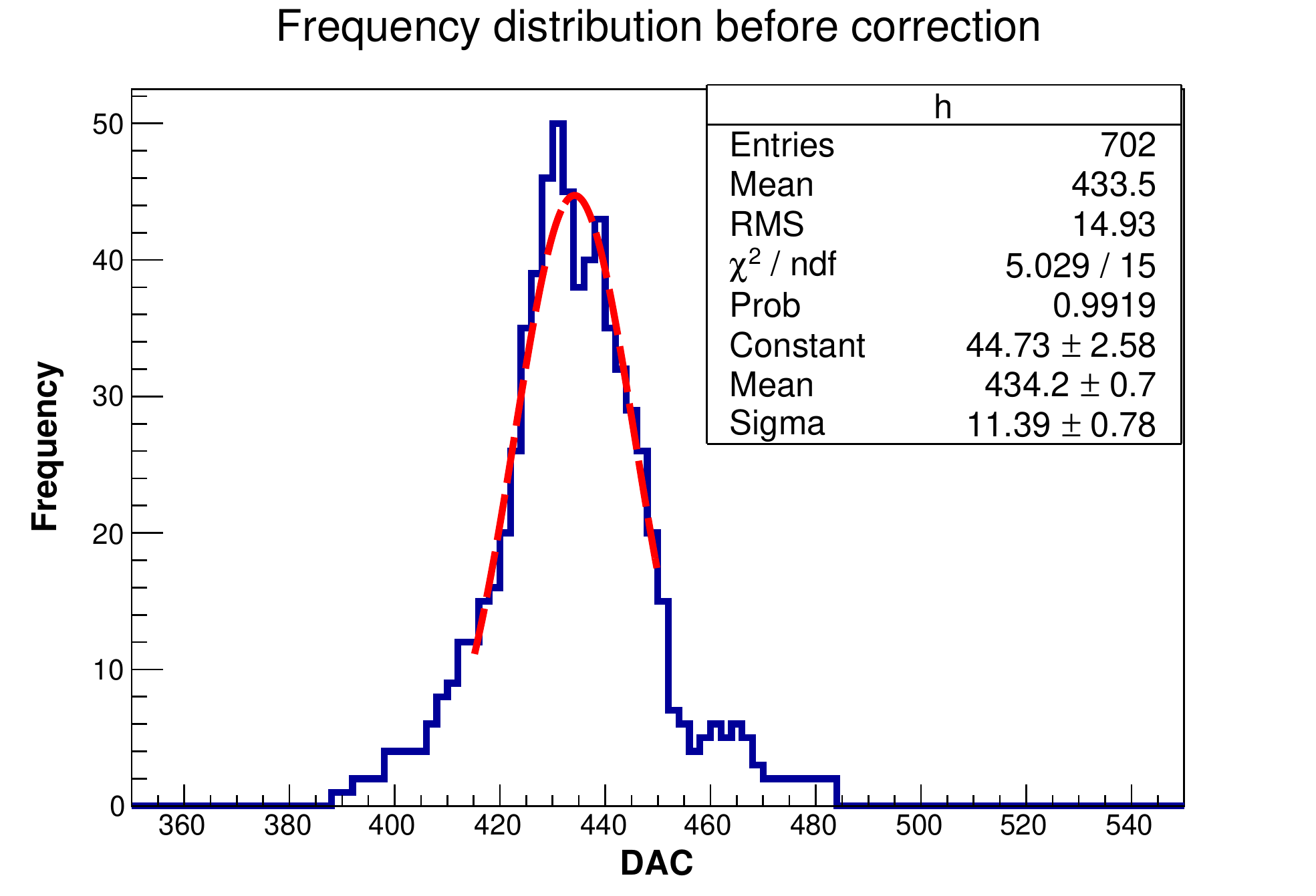}}
\subfloat[]{\includegraphics[width=0.5\linewidth,height=5cm]{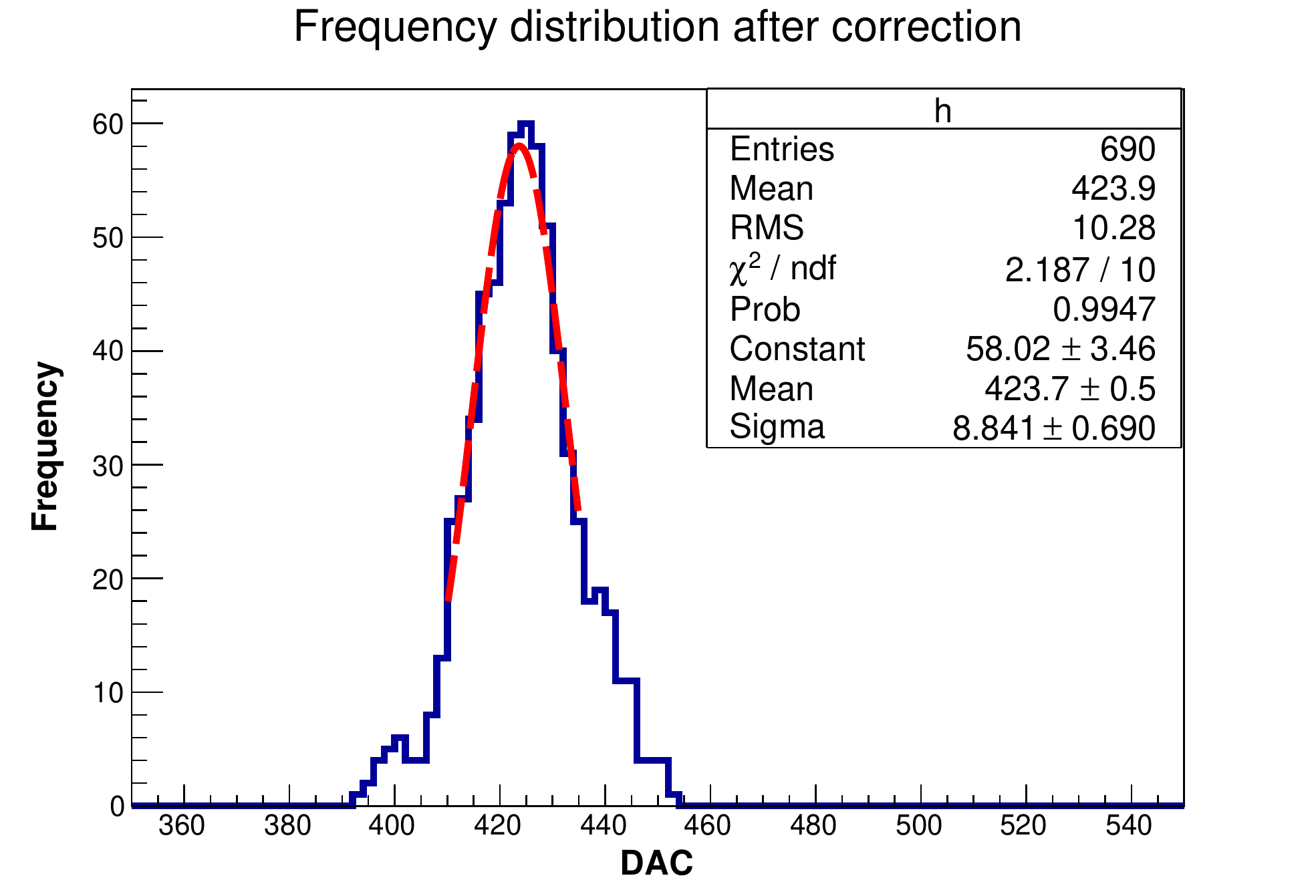}}
\caption{Distribution of the 64 inflection points (1 board with 1 ASIC) before (left) and after correction (right) with input charge of 100 fC.}
\label{gaus}
\end{figure}

For each channel the gain was chosen according to the aforementioned study so that the new inflection  value was as close as possible to the reference value. Once all the gains are corrected the S-curves were re-produced. Figure~\ref{gaus} shows the  50$\%$ efficiency of the inflection points obtained before and after the gain correction (as explained in next section). After the gain correction the dispersion of the 64 channels is reduced. To determine the correlation between the DAC unit and the injected charge, charges with different values were injected as shown in figure~\ref{ped}. It is found to be linear and fit with a straight line. The operation was repeated for all the 64 channels of the ASIC. The average value of the 64 inflection points obtained with gain = 1 was then selected as a reference value. The inflection point value determined at gain = 1. This allows one to find the correspondence between the charge and the DAC units which appears to be 1 fC = 1.22 DAC units. 

\begin{figure}[htbp]
\subfloat[]{\includegraphics[width=0.5\linewidth,height=5cm]{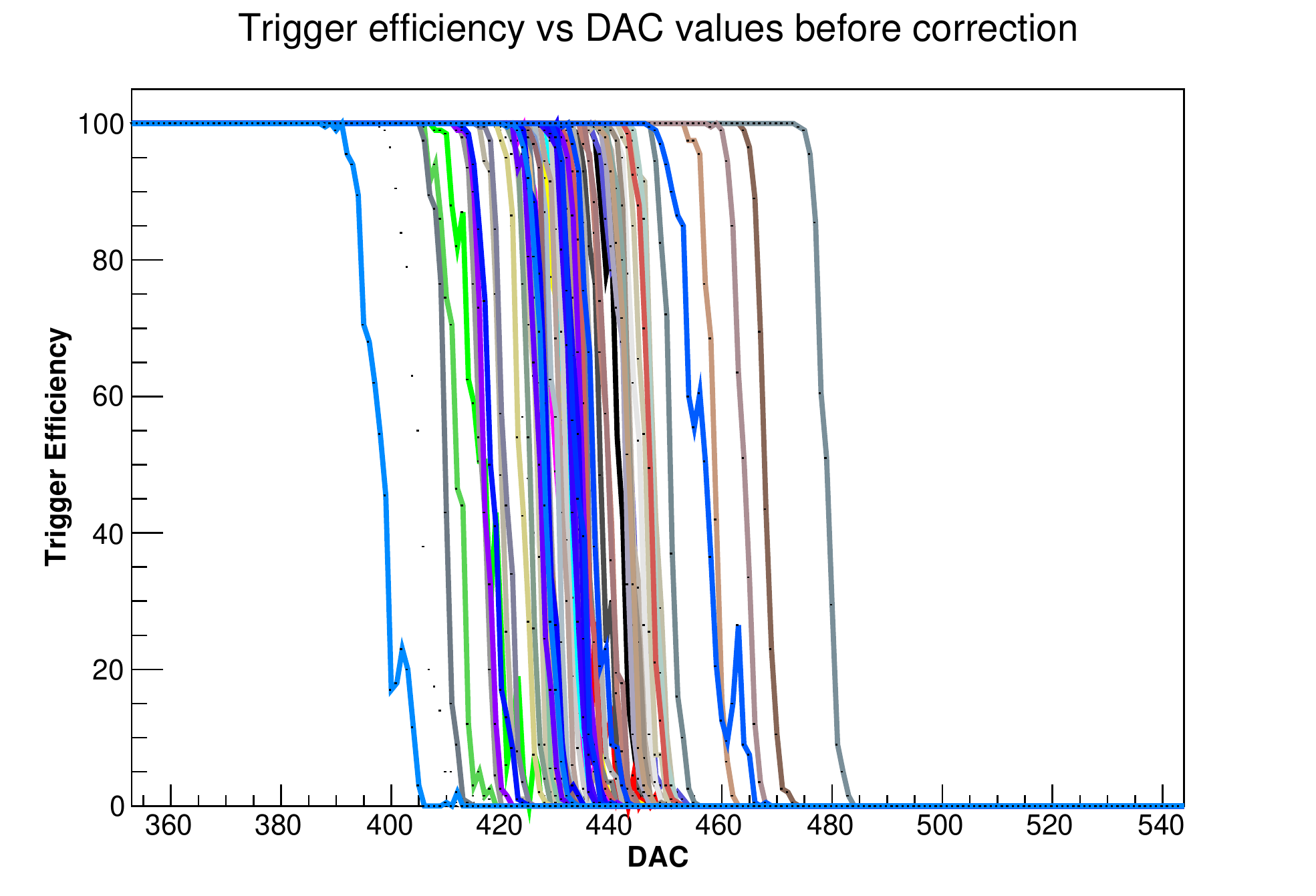}}
\subfloat[]{\includegraphics[width=0.5\linewidth,height=5cm]{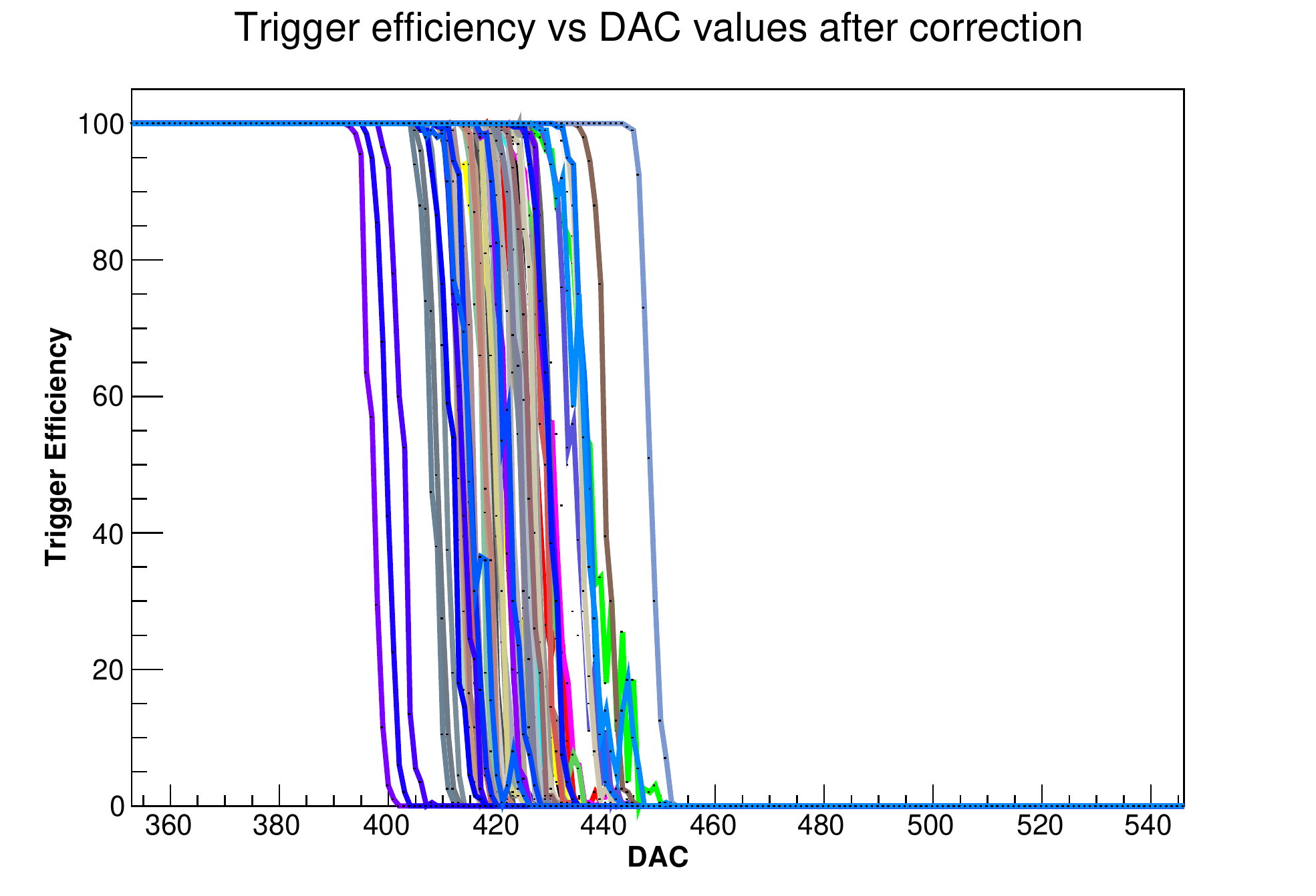}}
\caption{50$\%$ efficiency curve before (left) and after correction (right) with 100 fC charge input for all the 64 channels of the HARDROC ASIC.}
\label{trig}
\end{figure}

\subsection{Gain Correction Procedure}
An important feature of this ASIC is the possibility to change the gain of each channel in the interval 0 to 2 with a 8-bit precision. This is a useful tool to render the response of the different channels to the same signal as identical as possible. We started by injecting a given charge on each of the 64 channels using an arbitrary function generator (AFG). The threshold is varied over the whole dynamic range (1024) in steps of 1 DAC unit. The injection is repeated many times for each step. This allows the channel response efficiency to be estimated in terms of the applied threshold. The procedure is applied for different gain values. The inflection point on the efficiency curve for each channel and each gain is determined. The inflection point versus gain curves allow individual detector gains to be chosen to minimize the dispersion of the channel response for a given charge. The result of this procedure can be seen in figure~\ref{trig}, where the gain correction was applied for a 100 fC reference charge. The dispersion of the channels response after gain correction is lowered significantly. The small offset of the mean value is due to the constraint in adjusting gain settings. 

\section{Results}
The HARDROC is a semi-digital readout with three thresholds which allows excellent tracking capabilities with integrated data storage. There are three variable CRRC fast shapers with peaking time between 20-25 ns followed by three discriminators. The chip is so tiny that it can easily be embedded on the detector itself. It has the advantage to be connected to other chips by daisy chain mechanism so that chips of the same detector can communicate with each other. Slow control parameters as well as data collected in the detector can then circulate among the chips while the communication with the outside acquisition system is reduced significantly. In addition to these features, the chip has an internal memory capable of storing 128 events. After the calibration we performed various tests including trigger efficiency measurements before and after gain correction. The fluctuations between channels in terms of standard deviation ($\sigma$) can be corrected from 11.4 to 8.8. The 50$\%$ trigger efficiency measurement as a function of the injected charge also infer that the threshold can be set down to 5 fC which corresponds to the 5$\sigma$ noise limit. 

\section{Conclusions}
The design and measured performance of the HARDROC ASIC is presented keeping in view its immediate usage in the ICAL experiment. The proposed model of readout electronics can be easily embedded in the detector to eliminate many instrumentation issues. The HARDROC can accommodate 64 channels altogether in much smaller dimensions. The calibration of the ASIC has been performed using signals of various strengths. All the tests performed on the ASIC have shown good performance in efficiency measurements. The non-uniformities in the amplifier response are reduced significantly by controlling individual 8-bit precision settings. The linearity in response is tested for all three shapers. The standalone capabilities of the chip, which are crucial for its operation in INO-ICAL, are being tested and validated through test-bench measurements.


\acknowledgments

We acknowledge the financial support received from the Department of Science and Technology (DST), India and University of Delhi.

\end{document}